\documentstyle[12pt]{article}

\def\be{\begin{equation}}
\def\ee{\end{equation}}
\def\bea{\begin{eqnarray}}
\def\eea{\end{eqnarray}}
\def\ba{\begin{array}}
\def\ea{\end{array}}
\def\ben{\begin{enumerate}}
\def\een{\end{enumerate}}

\def\tx{\tilde{x}}

\def\lll{\label}
 
\begin{document}
\newcommand{\half}{{\textstyle\frac{1}{2}}}
\newcommand{\eqn}[1]{(\ref{#1})}
\newcommand{\npb}[3]{ {\bf Nucl. Phys. B}{#1} ({#2}) {#3}}
\newcommand{\pr}[3]{ {\bf Phys. Rep. }{#1} ({#2}) {#3}}
\newcommand{\prl}[3]{ {\bf Phys. Rev. Lett. }{#1} ({#2}) {#3}}
\newcommand{\plb}[3]{ {\bf Phys. Lett. B}{#1} ({#2}) {#3}}
\newcommand{\prd}[3]{ {\bf Phys. Rev. D}{#1} ({#2}) {#3}}
\newcommand{\hepth}[1]{ [{\bf hep-th}/{#1}]}
\newcommand{\grqc}[1]{ [{\bf gr-qc}/{#1}]}
 
\def\a{\alpha}
\def\b{\beta}
\def\g{\gamma}\def\G{\Gamma}
\def\d{\delta}\def\D{\Delta}
\def\ep{\epsilon}
\def\et{\eta}
\def\z{\zeta}
\def\t{\theta}\def\T{\Theta}
\def\l{\lambda}\def\L{\Lambda}
\def\m{\mu}
\def\f{\phi}\def\F{\Phi}
\def\n{\nu}
\def\p{\psi}\def\P{\Psi}
\def\r{\rho}
\def\s{\sigma}\def\S{\Sigma}
\def\ta{\tau}
\def\x{\chi}
\def\o{\omega}\def\O{\Omega}
\def\k{\kappa}
\def\pa {\partial}
\def\ov{\over}
\def\br{\nonumber\\}
\def\ud{\underline}

%%%%%%%%%%%%%%%%%%%%%%%%%%%%%%
\begin{flushright}
SINP/TNP/00-29\\
hep-th/0011236\\
\end{flushright}
\bigskip\bigskip
\begin{center}
{\Large\bf 
(NS5, D$p$) and (NS5, D$(p+2)$, D$p$) bound states of\\
type IIB and type IIA string theories}
\vskip1cm
{\sc  
Indranil Mitra and
Shibaji Roy}
\vskip0.5cm
Theory Division, Saha Institute of Nuclear Physics,\\
1/AF Bidhannagar, Calcutta-700 064, India 
\vskip 0.5cm
E-mails: indranil, roy@tnp.saha.ernet.in
\end{center}
\bigskip
\centerline{\bf ABSTRACT}
\bigskip

\begin{quote}
  
Starting from the $(q, p)$ 5-brane solution of type IIB string theory, we
here construct the low energy configuration corresponding to (NS5, D$p$)-brane
bound states (for $0\leq p \leq 4$) using the T-duality map between type IIB
and type IIA string theories. We use the SL(2, {\bf Z}) symmetry on the type 
IIB bound state (NS5, D3) to construct (NS5, D5, D3) bound state. We then 
apply T-duality transformation again on this state to construct the bound
states of the form (NS5, D$(p+2)$, D$p$) (for $0\leq p \leq 2$) of both type
IIB and type IIA string theories. We give the tension formula for these states
and show that they form non-threshold bound states. All these states preserve
half of the space-time supersymmetries of string theories. We also briefly
discuss the OD$p$-limits corresponding to (NS5, D$p$) bound state solutions.
\end{quote} 

\newpage
\section{Introduction}

The low energy effective actions of type II string theories are known to
possess an NS5-brane solution \cite{stro,chs,srey,dlu,dklu} which is the 
magnetic dual to the fundamental
string solution \cite{dh,dhr}. This solution is non-singular and 
purely solitonic and
therefore, not much is known about its dynamics. Type II string theories
also contain D$p$-branes \cite{hs,pol} in their low energy spectrum and 
it is well-known
that D$p$-branes can end on type IIA NS5-branes for $p$ = even and they can
end on type IIB NS5-branes for $p$ = odd. It is therefore expected that these
D$p$-branes will form bound states with NS5-branes. The bound state (NS5, D5)
of type IIB theory known as ($q$, $p$) fivebranes, with $q$, $p$ relatively 
prime integers corresponding to the charges of NS5-branes and D5-branes
respectively has already been constructed in ref.\cite{lurone}. We here use 
this solution
and apply the T-duality map from type IIB to type IIA theory also from type
IIA to type IIB theory to construct the (NS5, D$p$) bound state solutions
for $0 \leq p \leq 4$. The T-duality is applied along the longitudinal 
directions of D5-branes. We note that the (NS5, D$p$) bound states have also
been given in \cite{alior,alioz} and they were obtained from (NS5, D1) 
solution 
(constructed by applying S-duality on already known \cite{lurtwo} (F, D5) 
solution)
and applying T-duality along the transverse directions of D-string. However,
the (NS5, D5) solution obtained in this way does not agree with the already
known solution. We have indicated the possible reason for this discrepancy
in section 2. This is the reason we have chosen to start from the known 
(NS5, D5) solution to construct the (NS5, D$p$) solutions.

In order to construct (NS5, D$(p+2)$, D$p$) bound states for $0\leq p \leq 3$,
we start from (NS5, D3) solution of type IIB string theory. Since type IIB
string theory is conjectured to have a non-perturbative quantum SL(2, {\bf Z})
symmetry, we use this symmetry to construct (NS5, D5, D3) bound state first.
Then we apply T-duality map from type IIB (IIA) to type IIA (IIB) theory
along the longitudinal directions of D3-branes to construct (NS5, D$(p+2)$,
D$p$) bound states. We derive the tension formula for both (NS5, D$p$) and
(NS5, D$(p+2)$, D$p$) bound state solutions and show that they form 
non-threshold bound states. For (NS5, D$p$) case they are characterized by 
two relatively prime integers and for (NS5, D$(p+2)$, D$p$) case they are
characterized by three integers where any two of them are relatively prime.
Since S- and T-duality do not break supersymmetry of the system, all these
states preserve half of the space-time supersymmetries as the original 
(NS5, D5) solution of type IIB string theory.

Apart from being interesting on their own, one of the motivations for studying
the bound states of type II NS5-branes with D$p$-branes is to look at the
world-volume theory of NS5-branes in the presence of various D$p$-branes. It
is well-known that the world-volume theory of pure NS5-branes 
in the decoupling limit does not produce a local field theory. But gives a
non-gravitational, non-local theory known as the little string theory in 
(5+1)-dimensions \cite{bersei,sei,dvv1,dvv2,losms,oaha}. It is therefore, 
natural to look at the corresponding theory
in the presence of various D$p$-branes since they form bound states with
NS5-branes\footnote{Supergravity solutions involving NS5-branes in the presence
of RR fields have also been discussed in \cite{aliozs}.}. It is pointed out 
in ref.\cite{gopmss} that NS5-branes in the presence of
a critical RR electric field (D$p$-branes) reduce to again a non-gravitational
and non-local theory (different from little string theory) known as the
(5+1)-dimensional light open D$p$-brane (OD$p$) theories in a particular
decoupling limit. The (NS5, D$p$) bound state solutions in the decoupling
limit give the supergravity dual of OD$p$-theories. These supergravity 
solutions of OD$p$ theories are also discussed in ref.\cite{alior}. Since 
our solutions
differ from those given in \cite{alior}, we also briefly discuss the 
OD$p$-limits
for the solutions constructed in this paper. We have not studied the 
decoupling limits for (NS5, D$(p+2)$, D$p$) bound states in this paper, 
but it would be interesting to investigate what kind theories do they 
correspond to and will be reported elsewhere.

The paper is organized as follows. In section 2, we give the construction
of (NS5, D$p$) bound state solution starting from ($q$, $p$) 5-branes by
applying T-duality map along the longitudinal directions of D5-branes. We
also give the tension formula for these bound states. In section 3, starting
from the type IIB bound state (NS5, D3), we construct the (NS5, D5, D3)
bound state solution by applying SL(2, {\bf Z}) symmetry of type IIB theory.
In section 4, we construct the (NS5, D$(p+2)$, D$p$)-brane bound states by
applying T-duality on (NS5, D5, D3) along the longitudinal directions of 
D3-branes. Here also we give the corresponding tension formula. Finally, 
in section 5, we briefly discuss the OD$p$ limits for the (NS5, D$p$) solutions
obtained in section 2. 

\section{(NS5, D$p$) bound states}

Type IIB string theory is well-known to possess an SL(2, {\bf Z}) multiplet
of magnetically charged 5-brane solution known as $(q, p)$ 5-branes first
constructed in ref.\cite{lurone}. In this section we will start with this 
solution and
apply T-duality map between type IIB and type IIA string theories along the 
longitudinal directions of 5-branes to construct the (NS5, D$p$) bound state
solutions. The $(q, p)$ 5-brane solution, with $(q, p)$ relatively prime
integers, denoting the charges of NS5-brane and D5-brane respectively is given
as,

\noindent{\bf \underline{(NS5, D5) solution:}}
\bea
ds^2 &=& H^{1/2} H'^{1/2}\left[H^{-1}\left(-dx_0^2 + dx_1^2 + \cdots +
dx_5^2\right) + dr^2 + r^2 d\Omega_3^2\right]\br
e^{\phi_b} &=& g_s H' H^{-1/2}\br
B^{(b)} &=& 2 Q_5 \cos\varphi \sin^2\theta \cos\phi_1 d\theta \wedge d\phi_2\br
A^{(2)} &=& - \frac{2 Q_5 \sin\varphi}{g_s} \sin^2\theta \cos\phi_1 d\theta
\wedge d\phi_2\br
\chi &=& -\frac{p}{q} (1 - H'^{-1}), \qquad\qquad A^{(4)} = 0
\lll{1}
\eea
where in the above, we have written the metric in the string-frame. Also,
$d\Omega_3^2 = d\theta^2 + \sin^2\theta d\phi_1^2 + \sin^2\theta \sin^2\phi_1
d\phi_2^2$ is the line element for the unit 3-sphere transverse to the
5-branes, $g_s = e^{\phi_0}$ is the asymptotic value of the dilaton, $B^{(b)}$
and $A^{(2)}$ denote the NSNS and RR two-form potentials. $\chi$ is the RR
scalar and $A^{(4)}$ is the RR 4-form gauge field. The harmonic functions
$H$ and $H'$ are given as,
\bea
H &=& 1 + \frac{Q_5}{r^2}\br
H' &=& 1 + \frac{\cos^2\varphi Q_5}{r^2}
\lll{2}
\eea
where the angle $\cos\varphi = \frac{q}{\sqrt{p^2 g_s^2 + q^2}}$ and $Q_5$
is defined as,
\be
Q_5 = \sqrt{p^2 g_s^2 + q^2}\,\alpha'
\lll{3}
\ee
Note that the solution given here is different from the one constructed in
\cite{alior,alioz}. The axion in this case vanishes asymptotically, whereas
it is constant in \cite{alioz}. Also, the RR 3-form field strength is constant
here, but in \cite{alioz} it is proportional to $\chi$. The other fields are
the same.

The T-duality map from type IIB fields to type IIA fields are given as 
\cite{berho},
\bea
G_{\tx \tx} &=& \frac{1}{J_{\tx \tx}}, \qquad\qquad G_{\tx \mu} = - \frac{
B_{\tx \mu}^{(b)}}{J_{\tx \tx}}\br
G_{\mu\nu} &=& J_{\mu\nu} - \frac{J_{\tx \mu} J_{\tx \nu} - B_{\tx \mu}^{(b)}
B_{\tx \nu}^{(b)}}{J_{\tx \tx}}\br
e^{2\phi_a} &=& \frac{e^{2\phi_b}}{J_{\tx\tx}}\br
B_{\tx \mu}^{(a)} &=& - \frac{J_{\tx \mu}}{J_{\tx\tx}}, \qquad\qquad
B_{\mu\nu}^{(a)} = B_{\mu\nu}^{(b)} + 2 \frac{B_{\tx [\mu}^{(b)} J_{\nu] \tx}}
{J_{\tx\tx}}\br
A_{\tx}^{(1)} &=& -\chi, \qquad\qquad A_\mu^{(1)} = A_{\tx \mu}^{(2)} +
\chi B_{\tx \mu}^{(b)}\br
A_{\tx \mu\nu}^{(3)} &=& A_{\mu\nu}^{(2)} + 2\frac{A_{\tx [\mu}^{(2)}
J_{\nu]\tx}}{J_{\tx\tx}}\br
A_{\mu\nu\rho}^{(3)} &=& A_{\mu\nu\rho\tx}^{(4)} + \frac{3}{2}\left(
A_{\tx[\mu}^{(2)} B_{\nu\rho]}^{(b)} - B_{\tx[\mu}^{(b)} A_{\nu\rho]}^{(2)}
- 4 \frac{B_{\tx[\mu}^{(b)} A^{(2)}_{\tx\nu} J_{\rho]\tx}}{J_{\tx\tx}}\right)
\lll{4}
\eea
Here $\tx$ is the Killing coordinate along which T-duality is performed. $\mu,\,
\nu,\,\rho$ denote any coordinate other than $\tx$. $J$ and $G$ are the 
string-frame metric in type IIB and type IIA string theory respectively. 
$\phi_a$,
$\phi_b$ are the dilatons and $B^{(b)}$, $B^{(a)}$ are the NSNS two-form
gauge fields in these two theories. $A^{(1)}$ and $A^{(3)}$ are the RR 1-form
and 3-form gauge fields in type IIA theory whereas, $\chi$, $A^{(2)}$ and
$A^{(4)}$ are the RR scalar, 2-form and 4-form gauge fields in type IIB 
string theory. The field strengths which appear in the corresponding low energy
actions are given as,
\bea
H^{(a)} &=& d B^{(a)}\br
H^{(b)} &=& d B^{(b)}\br
F^{(2)} &=& d A^{(1)}\br
F^{(3)} &=& d A^{(2)}\br
F^{(4)} &=& d A^{(3)} - H^{(a)} \wedge A^{(1)}\br
F^{(5)} &=& d A^{(4)} - \frac{1}{2}\left(B^{(b)} \wedge F^{(3)} - A^{(2)}
\wedge H^{(b)}\right)
\lll{5}
\eea
Note that the RR 5-form field strength $F^{(5)}$ in type IIB theory is 
self-dual i.e. $F^{(5)} = \ast F^{(5)}$, with $\ast$ denoting the Hodge-dual
and we will use this fact to find out the gauge field $A^{(4)}$ in (NS5, D3)
solution in the following.

Now a straightforward application of the T-duality rule given in eq.\eqn{4}
on the type IIB background (1) along $x^5$ coordinate yields the (NS5, D4)
bound state configuration in type IIA theory as,

\noindent{\bf \underline{(NS5, D4) solution:}}
\bea
ds^2 &=& H^{1/2} H'^{1/2}\left[H^{-1}\left(-dx_0^2 + dx_1^2 + \cdots +
dx_4^2\right) + H'^{-1} dx_5^2 + dr^2 + r^2 d\Omega_3^2\right]\br
e^{\phi_a} &=& g_s H'^{3/4} H^{-1/4}\br
B^{(a)} &=& 2 Q_5 \cos\varphi \sin^2\theta \cos\phi_1 d\theta \wedge d\phi_2\br
A^{(1)} &=& \frac{p}{q}(1-H'^{-1}) dx^5\br
A^{(3)} &=& - \frac{2 Q_5 \sin\varphi}{g_s} \sin^2\theta \cos\phi_1 dx^5
\wedge d\theta
\wedge d\phi_2\br
\lll{6}
\eea
Now in order to get (NS5, D3) solution from here, we have to apply T-duality
transformation along $x^4$-direction. Since the solution (6) belongs to
type IIA theory, we have to use the T-duality map from type IIA to type IIB
fields. They are given as follows \cite{berho},
\bea
J_{\tx \tx} &=& \frac{1}{G_{\tx \tx}}, \qquad\qquad J_{\tx \mu} = - \frac{
B_{\tx \mu}^{(a)}}{G_{\tx \tx}}\br
J_{\mu\nu} &=& G_{\mu\nu} - \frac{G_{\tx \mu} G_{\tx \nu} - B_{\tx \mu}^{(a)}
B_{\tx \nu}^{(a)}}{G_{\tx \tx}}\br
e^{2\phi_b} &=& \frac{e^{2\phi_a}}{G_{\tx\tx}}\br
B_{\tx \mu}^{(b)} &=& - \frac{G_{\tx \mu}}{G_{\tx\tx}}, \qquad\qquad
B_{\mu\nu}^{(b)} = B_{\mu\nu}^{(a)} + 2 \frac{G_{\tx [\mu} B_{\nu] \tx}^{(a)}}
{G_{\tx\tx}}\br
\chi &=& -A_{\tx}^{(1)}, \qquad\qquad A_{\tx \mu}^{(2)} = A_\mu^{(1)} -
\frac{A_{\tx}^{(1)} G_{\tx \mu}}{G_{\tx\tx}}\br
A_{\mu\nu}^{(2)} &=& A_{\mu\nu\tx}^{(3)} - 2A_{[\mu}^{(1)}B_{\nu]\tx}^{(a)}+ 
2\frac{G_{\tx[\mu} B_{\nu]\tx}^{(a)} A_{\tx}^{(1)}}
{G_{\tx\tx}}\br
A_{\mu\nu\rho\tx}^{(4)} &=& A_{\mu\nu\rho}^{(3)} - \frac{3}{2}\left(
A_{[\mu}^{(1)} B_{\nu\rho]}^{(a)} - \frac{G_{\tx[\mu} B_{\nu\rho]}^{(a)}
A_{\tx}^{(1)}}{G_{\tx\tx}}
+  \frac{G_{\tx[\mu} A_{\nu\rho]\tx}^{(3)}}{G_{\tx\tx}}\right)
\lll{7}
\eea
Here as before, $\tx$ is the Killing coordinate along which T-duality 
transformation is performed and $\mu,\,\nu,\,\rho$ are any coordinate
other than $\tx$. Also, the remaining components of $A^{(4)}$ can be obtained 
from the self-duality condition on the corresponding 5-form field-strength.
Thus applying T-duality transformation along $x^4$-coordinate we obtain
(NS5, D3) solution of type IIB theory as follows:

\noindent{\bf \underline{(NS5, D3) solution:}}
\bea
ds^2 &=& H^{1/2} H'^{1/2}\left[H^{-1}\left(-dx_0^2 + dx_1^2 + \cdots +
dx_3^2\right) + H'^{-1} \left(dx_4^2 + dx_5^2\right) + 
dr^2 + r^2 d\Omega_3^2\right]\br
e^{\phi_b} &=& g_s H'^{1/2} \br
B^{(b)} &=& 2 Q_5 \cos\varphi \sin^2\theta \cos\phi_1 d\theta \wedge d\phi_2\br
\chi &=& 0\br
A^{(2)} &=& \frac{p}{q}(1-H'^{-1}) dx^4 \wedge dx^5\br
A^{(4)} &=& - \frac{Q_5 \sin\varphi}{g_s}(1+H'^{-1})\sin^2\theta \cos\phi_1 
dx^4 \wedge dx^5 \wedge d\theta \wedge d\phi_2\br 
&& \qquad - \frac{\sin\varphi}{g_s}
H^{-1} dx^0 \wedge dx^1 \wedge dx^2 \wedge dx^3
\lll{8}
\eea
We would like to point out that the T-duality rule given in \eqn{7} produces
only the first term of $A^{(4)}$ in eq.\eqn{8}. However, since D3-branes are
self-dual, we have to include the term obtained by applying self-duality on
the corresponding field strength. This how we obtained the second term of
$A^{(4)}$ above. This is also important to obtain the correct form of (NS5, D4)
and (NS5, D5) solution if we apply T-duality in the transverse directions of
D3-branes in (NS5, D3) solution. We think this is the reason why (NS5, D5)
solution obtained in \cite{alioz} differs from eq.\eqn{1}.

We now apply T-duality map from type IIB fields to type IIA fields given
in eq.\eqn{4} along $x^3$-coordinate on (NS5, D3) solution to obtain (NS5, D2)
bound state configuration,

\noindent{\bf \underline{(NS5, D2) solution:}}
\bea
ds^2 &=& H^{1/2} H'^{1/2}\left[H^{-1}\left(-dx_0^2 + dx_1^2 + dx_2^2\right) 
+ H'^{-1} \left(dx_3^2 + dx_4^2 + dx_5^2\right) + 
dr^2 + r^2 d\Omega_3^2\right]\br
e^{\phi_a} &=& g_s H^{1/4} H'^{1/4} \br
B^{(a)} &=& 2 Q_5 \cos\varphi \sin^2\theta \cos\phi_1 d\theta \wedge d\phi_2\br
A^{(1)} &=& 0\br
A^{(3)} &=& \frac{p}{q}(1-H'^{-1}) dx^3 \wedge dx^4 \wedge dx^5 -
\frac{\sin\varphi}{g_s} H^{-1} dx^0 \wedge dx^1 \wedge dx^2
\lll{9}
\eea
This state belongs to type IIA theory. So applying T-duality map from type IIA
to type IIB fields given in (7) along $x^2$-coordinate, we obtain (NS5, D1)
bound state as follows:

\noindent{\bf \underline{(NS5, D1) solution:}}
\bea
ds^2 &=& H^{1/2} H'^{1/2}\left[H^{-1}\left(-dx_0^2 + dx_1^2\right) 
+ H'^{-1} \left(dx_2^2 + \cdots + dx_5^2\right) + 
dr^2 + r^2 d\Omega_3^2\right]\br
e^{\phi_b} &=& g_s H^{1/2} \br
B^{(b)} &=& 2 Q_5 \cos\varphi \sin^2\theta \cos\phi_1 d\theta \wedge d\phi_2\br
\chi &=& 0\br
A^{(2)} &=& 
-\frac{\sin\varphi}{g_s} H^{-1} dx^0 \wedge dx^1\br
A^{(4)} &=& -\frac{p}{q}(1-H'^{-1}) dx^2 \wedge dx^3 \wedge dx^4 \wedge dx^5
\lll{10}
\eea
Finally the (NS5, D0) bound state can be obtained from the above type IIB
bound state configuration by applying T-duality map eq.\eqn{4} along 
$x^1$-direction. The solution is:

\noindent{\bf \underline{(NS5, D0) solution:}}
\bea
ds^2 &=& H^{1/2} H'^{1/2}\left[-H^{-1}dx_0^2  
+ H'^{-1} \left(dx_1^2 + \cdots + dx_5^2\right) + 
dr^2 + r^2 d\Omega_3^2\right]\br
e^{\phi_a} &=& g_s H^{3/4} H'^{-1/4} \br
B^{(a)} &=& 2 Q_5 \cos\varphi \sin^2\theta \cos\phi_1 d\theta \wedge d\phi_2\br
A^{(1)} &=& \frac{\sin\varphi}{g_s} H^{-1} dx^0\br
A^{(3)} &=& 
\frac{Q_5\sin\varphi\cos\varphi}{g_s} H^{-1}\sin^2\theta\cos\phi_1 
dx^0 \wedge d\theta \wedge d\phi_2\br
\lll{11}
\eea
Note here that the (NS5, D0), (NS5, D1) and (NS5, D2) solutions given above 
match precisely with the solutions in \cite{alior} apart from some unimportant
constant term in RR gauge fields.

The ADM mass as well as the tension for these (NS5, D$p$) bound states can be
calculated by a generalization of the mass formula given in \cite{lu}. This
has been done in \cite{lurtwo} for (F, D$p$) solutions. Here we use the same 
technique to obtain the tension of (NS5, D$p$) bound states as
\be
T_{(q,p)} = \frac{1}{g_s^2}\sqrt{p^2 g_s^2 + q^2}\,\frac{1}{(2\pi)^5 \alpha'^3}
\lll{t1}
\ee
Note that the tension is proportional to the charge $Q_5$ given in \eqn{3}.
Here $1/[(2\pi)^5\alpha'^3]$ is the fundamental tension of a 5-brane. For
a single NS5-brane $q=1$, $p=0$, we recover the tension of an NS5-brane as
$1/[g_s^2(2\pi)^5\alpha'^3]$. On the other hand for D5 brane $q=0$ and $p=1$,
we recover the tension of a D5-brane as $1/[g_s(2\pi)^5\alpha'^3]$. For other
D$p$-branes with $p<5$, there are infinite number D$p$-branes in the
world-volume of NS5-brane lying along $p$ spatial directions. `$p$' in the
expression\footnote{Although we have denoted the charge of a D$p$-brane
by the same integer `$p$', they should not be confused and should be clear 
from the context.} \eqn{t1} denotes the charge of  D$p$-branes per 
$(2\pi)^{5-p} \alpha'^{(5-p)/2}$ of $(5-p)$-dimensional area of NS5-brane.
So, for example, the tension of a D-string in (NS5, D1) bound state obtained
from \eqn{t1} is given by $\frac{1}{g_s}\frac{p}{(2\pi)^5\alpha'^3} (2\pi)^4
\alpha'^2 = \frac{p}{g_s(2\pi\alpha')}$ as expected. It is clear from the
expression \eqn{t1} that when $q$, $p$ are relatively prime integers (NS5,
D$p$) form non-threshold bound states.

\section{SL(2, {\bf Z}) transformation and (NS5, D5, D3) bound state
solution}

We have obtained the type IIB bound state (NS5, D3) in the previous section.
Since type IIB string theory is well-known to possess a non-perturbative
quantum SL(2, {\bf Z}) symmetry, we will use it in this section on (NS5, D3)
solution to construct (NS5, D5, D3) bound state solution. Note that (NS5, D1)
solution also constructed in the previous section belongs to type IIB theory
as well and we could use SL(2, {\bf Z}) symmetry on this state to construct
((NS5, D5), (D1, F)) bound state, but this has already been done in 
\cite{lurthree}.
So, (NS5, D5, D3) is the only new solution involving two D-branes in this
case and we will construct it in this section following 
refs.\cite{lurone,lurfour}. We here write 
the (NS5, D3) solution with $g_s = 1$, since this $g_s$ has nothing to do
with the asymptotic value of the dilaton in the final (NS5, D5, D3) bound
state. (We will, however, restore the string coupling constant when we finally
construct this bound state).
\bea
ds^2 &=& H^{1/2} H'^{1/2}\left[H^{-1}\left(-dx_0^2 + dx_1^2 + \cdots +
dx_3^2\right) + H'^{-1} \left(dx_4^2 + dx_5^2\right) + 
dr^2 + r^2 d\Omega_3^2\right]\br
e^{\phi_b} &=& H'^{1/2} \br
B^{(b)} &=& 2 Q_5 \cos\varphi \sin^2\theta \cos\phi_1 d\theta \wedge d\phi_2\br
\chi &=& 0\br
A^{(2)} &=& \frac{p}{q}(1-H'^{-1}) dx^4 \wedge dx^5\br
A^{(4)} &=& - Q_5 \sin\varphi(1+H'^{-1})\sin^2\theta \cos\phi_1 
dx^4 \wedge dx^5 \wedge d\theta \wedge d\phi_2\br 
&& \qquad - \sin\varphi
H^{-1} dx^0 \wedge dx^1 \wedge dx^2 \wedge dx^3
\lll{12}
\eea
where
\bea
H &=& 1 + \frac{Q_5}{r^2}\br
H' &=& 1 + \frac{\cos^2\varphi Q_5}{r^2}
\lll{13}
\eea
with $Q_5 = \sqrt{p^2 + q^2} \alpha'$, $\cos\varphi = \frac{q}{\sqrt{p^2 +
q^2}}$. Here $p$ is the D3-brane charge and $q$ is the NS5-brane charge.
Also, the metric in the above is written in the string-frame and we have
to write in the Einstein-frame since the Einstein-frame metric remains
invariant under SL(2, {\bf Z}) transformation. We will first make a classical
SL(2, {\bf R}) transformation on (13) and then impose the charge quantization
to obtain the final (NS5, D5, D3) solution. The Einstein-frame metric has
the form
\bea
ds_E^2 &=& e^{-\phi_b/2} ds^2\br
&=& H^{1/2} H'^{1/4}\left[H^{-1}\left(-dx_0^2 + dx_1^2 + \cdots + dx_3^2\right)
+ H'^{-1}\left(dx_4^2 + dx_5^2\right) + dr^2 + r^2 d\Omega_3^2\right]
\lll{14}
\eea
If $\Lambda$ denotes the global SL(2, {\bf R}) transformation matrix then
type IIB fields transform under this transformation as follows:
\bea
g_{\mu\nu}^E &\rightarrow& g_{\mu\nu}^E, \qquad \lambda \rightarrow
\frac{a\lambda + b}{c\lambda + d}\br
\left(\begin{array}{c} B^{(b)} \\ A^{(2)}\end{array}\right) &\rightarrow&
(\Lambda^T)^{-1}
\left(\begin{array}{c} B^{(b)} \\ A^{(2)}\end{array}\right), \qquad
\left(\begin{array}{c} Q_1 \\ Q_2\end{array}\right) \rightarrow
(\Lambda^T)^{-1} \left(\begin{array}{c} Q_1 \\ Q_2\end{array}\right)\br
A^{(4)} &\rightarrow& A^{(4)}, \qquad Q_3 \rightarrow Q_3
\lll{15}
\eea
where $\Lambda = \left(\begin{array}{cc} a & b\\ c & d\end{array}\right)$,
with $ad - bc = 1$. Also, `$T$' here denotes the transpose of a matrix.
$\lambda = \chi + i e^{-\phi}$ and $Q_1$ and $Q_2$ denote the charges of the
NS5-brane and D5-brane respectively. $Q_3$ is the D3-brane charge which remains
invariant under SL(2, {\bf Z}). Note here that since the 5-brane charges 
are topological (magnetic), the charges transform in the same way as the
gauge fields $B^{(b)}$ and $A^{(2)}$. If we assume the asymptotic value of
the axion to be zero\footnote{This is taken for simplicity. One can construct
the (NS5, D5, D3) solution with a non-zero asymptotic value of the axion 
($\chi_0$) by
further making an SL(2, {\bf R}) transformation on our solution given later
in eqs.(25)--(28) with the SL(2, {\bf R}) matrix $\left(\begin{array}{cc}
1 & \chi_0\\ 0 & 1\end{array}\right)$.} and the asymptotic value of the 
dilaton to be $\phi_0$,
then the SL(2, {\bf R}) matrix take the following form:
\be
\Lambda = \left(\begin{array}{cc} e^{-\phi_0/2} \cos\alpha &
- e^{-\phi_0/2}\sin\alpha\\e^{\phi_0/2}\sin\alpha & e^{\phi_0/2}\cos\alpha
\end{array}\right)
\lll{16}
\ee
where $\alpha$ is an unknown parameter to be determined from the charge
quantization condition. For the initial (NS5, D3) configuration $q$ was
the charge of NS5-brane and D5-brane charge was zero, where $q$ is an integer.
We replace the initial charge of NS5-brane by an unknown number $\Delta^{1/2}$
(which is no longer an integer) and then impose the charge quantization
after the SL(2, {\bf R}) transformation. So,
\be
\left(\begin{array}{c}m\\n\end{array}\right) = \left(\begin{array}{cc}
e^{\phi_0/2} \cos\alpha & -e^{\phi_0/2} \sin\alpha\\ e^{-\phi_0/2}\sin\alpha
& e^{-\phi_0/2}\cos\alpha\end{array}\right) \left(\begin{array}{c}
\Delta^{1/2}\\ 0 \end{array}\right)
\lll{17}
\ee
Here $(m, n)$ are integers and are associated with the charges of the NS5-brane
and D5-brane respectively in the final configuration. From (18) we obtain,
\bea
\sin\alpha &=& e^{\phi_0/2} \Delta^{-1/2} n\br
\cos\alpha &=& e^{-\phi_0/2} \Delta^{-1/2} m
\lll{18}
\eea
The above equation determines the value of $\Delta$ to be of the form
\be
\Delta = \left(m^2 e^{-\phi_0} + n^2 e^{\phi_0}\right)
\lll{19}
\ee
The SL(2, {\bf R}) matrix $\Lambda$ in (17) therefore takes the form
\be
\Lambda = \frac{1}{\sqrt{m^2 e^{-\phi_0} + n^2 e^{\phi_0}}}
\left(\begin{array}{cc} e^{-\phi_0} m & -n\\ e^{\phi_0} n & m\end{array}\right)
\lll{20}
\ee
With this form of the SL(2, {\bf R}) matrix, the dilaton and the axion for
the (NS5, D5, D3) bound state are:
\bea
e^{\phi_b} &=& e^{\phi_0} H'^{-1/2} H''\br
\chi &=& \frac{mn(1-H')}{H''(m^2 + n^2 g_s^2)}
\lll{21}
\eea
where $H$ and $H'$ are as given in \eqn{13} with $q$ replaced by $\Delta^{1/2}$
(given in \eqn{19}) and $H''$ is given as
\be
H'' = 1 + \frac{m^2 e^{-\phi_0}Q_5/(p^2 + m^2 e^{-\phi_0} + n^2 e^{\phi_0})}
{r^2}
\lll{22}
\ee
The metric retains its form as given in \eqn{14}, with the proper replacement 
of $q$ as mentioned above. The four-form gauge field also remains invariant and
takes the form as given in \eqn{12} whereas, the NSNS and RR two-form gauge
fields transform according to eq.\eqn{15} and the final forms are
\bea
B^{(b)} &=& \frac{2m}{\sqrt{m^2 e^{-\phi_0} + n^2 e^{\phi_0}}}Q_5\cos\varphi
\sin^2\theta \cos\phi_1 d\theta \wedge d\phi_2\br
&&\qquad -\frac{npe^{\phi_0}}{(m^2 e^{-\phi_0} + n^2 e^{\phi_0})}(1-H'^{-1})
dx^4 \wedge dx^5\br
A^{(2)} &=& \frac{2n}{\sqrt{m^2 e^{-\phi_0} + n^2 e^{\phi_0}}}Q_5\cos\varphi
\sin^2\theta \cos\phi_1 d\theta \wedge d\phi_2\br
&&\qquad +\frac{mpe^{-\phi_0}}{(m^2 e^{-\phi_0} + n^2 e^{\phi_0})}(1-H'^{-1})
dx^4 \wedge dx^5
\lll{23}
\eea
So, we have constructed the (NS5, D5, D3) bound state with the Einstein
metric in \eqn{14}, the dilaton and axion as given in \eqn{21}, the NSNS
and RR two-form gauge fields given in \eqn{23} and the SL(2, {\bf Z}) invariant
4-form gauge field given in \eqn{12}. Now we want to write this solution in the
string frame such that the string-frame metric has the asymptotically 
Minkowskian form. Note here that in the (NS5, D5, D3) solution we have 
constructed the Einstein frame metric (eq.\eqn{14}) is asymptotically 
Minkowskian.
If we naively convert it into string frame by multiplying it with 
$e^{\phi_b/2}$, with $e^{\phi_b}$ given in \eqn{21}, then the string 
frame metric
does not become asymptotically Minkowskian. However, this can be achieved
by scaling the coordinates as $(x_0,\,x_1,\cdots, x_5,\,r) \rightarrow
e^{-\phi_0/4}(x_0,\,x_1,\cdots, x_5,\,r)$. We would also like to point
out that in order to restore the correct $g_s$ dependence, we have to replace
$p$ by $e^{\phi_0/2} p$ everywhere. The reason for this is that the D5-brane
charge $n$ and D3-brane charge $p$ should have the same $g_s$ factors 
multiplied since their masses and the charges have the same $g_s$ dependence.
So, with the above rescaling of the coordinates and the above replacement of
the integer $p$, we can write down the (NS5, D5, D3) solution by the following
string frame metric, dilaton, axion and other gauge fields,
\bea
ds^2 &=& 
H^{1/2} H''^{1/2}\left[H^{-1}\left(-dx_0^2 + dx_1^2 + \cdots + dx_3^2\right)
+ H'^{-1}\left(dx_4^2 + dx_5^2\right) + dr^2 + r^2 d\Omega_3^2\right]\br
e^{\phi_b} &=& g_s H'^{-1/2} H''\br
\chi &=& \frac{mn(1-H')}{H''(m^2 + n^2 g_s^2)} = \frac{n}{m}(H''^{-1}-1)\br
B^{(b)} &=& \frac{2m}{\sqrt{m^2 + n^2 g_s^2}}Q_5\cos\varphi
\sin^2\theta \cos\phi_1 d\theta \wedge d\phi_2
-\frac{np g_s^2}{(m^2 + n^2 g_s^2)}(1-H'^{-1})
dx^4 \wedge dx^5\br
A^{(2)} &=& \frac{2n}{\sqrt{m^2 + n^2 g_s^2}}Q_5\cos\varphi
\sin^2\theta \cos\phi_1 d\theta \wedge d\phi_2
+\frac{mp}{(m^2 + n^2 g_s^2)}(1-H'^{-1})
dx^4 \wedge dx^5\br
A^{(4)} &=& -\frac{Q_5}{g_s} \sin\varphi (1+H'^{-1})\sin^2\theta \cos\phi_1
dx^4 \wedge dx^5 \wedge d\theta \wedge d\phi_2\br
&&\qquad -\frac{\sin\varphi}{g_s} H^{-1} dx^0 \wedge dx^1 \wedge dx^2 \wedge
dx^3
\lll{24}
\eea
Here $H$ and $H'$ are as given in \eqn{13} and
\bea
H'' &=& 1 + \frac{m^2 Q_5/(m^2 + (p^2 + n^2)g_s^2)}{r^2}\\
\cos\varphi &=& \frac{(m^2 + n^2 g_s^2)^{1/2}}{(m^2 + (p^2 + n^2)g_s^2)^{1/2}}
\lll{2526}
\eea
The form of $Q_5$ is
\be
Q_5 = \sqrt{m^2 + (p^2 + n^2)g_s^2} \alpha'
\lll{27}
\ee
Thus eqs.(25)--\eqn{27} represent the (NS5, D5, D3) solution with 
$m,\,n,\,p$
representing the integral charges for NS5-brane, D5-brane and D3-brane 
respectively. It can be easily checked that for $n=0$, the above solution 
reduces to (NS5, D3) solution given in eq.(8) and for $p=0$, it reduces to
(NS5, D5) solution given in \eqn{1}. Finally, for $m=0$, this solution reduces
to (D5, D3) solution constructed in ref.\cite{breckmm,cosp}.

\section{T-duality and (NS5, D$(p+2)$, D$p$) bound states}

The (NS5, D5, D3) solution constructed in the previous section belongs to
type IIB theory, where NS5 and D5 branes are lying along the spatial
$x^1,\,x^2,\,x^3,\,x^4,\,{\rm and}\,x^5$ directions. D3 branes lie along
$x^1,\,x^2,\,x^3$ directions. So, if we apply T-duality map from type IIB 
theory to type IIA theory given in eq.(4) along $x^3$ direction, then we will
get the (NS5, D4, D2) bound state solution with D4-branes lying along $x^1,\,
x^2,\,x^4,\,x^5$ coordinates and D2-branes lying along $x^1,\,x^2$ coordinates.
A straightforward application of T-duality map produces the following
(NS5, D4, D2) bound state solution:

\noindent{\bf \underline{(NS5, D4, D2) solution:}}
\bea
ds^2 &=& 
H^{1/2} H''^{1/2}\left[H^{-1}\left(-dx_0^2 + dx_1^2 + dx_3^2\right) + H''^{-1}
dx_3^2
+ H'^{-1}\left(dx_4^2 + dx_5^2\right) + dr^2 + r^2 d\Omega_3^2\right]\br
e^{\phi_a} &=& g_s H^{1/4}H'^{-1/2} H''^{3/4}\br
B^{(a)} &=& \frac{2m}{\sqrt{m^2 + n^2 g_s^2}}Q_5\cos\varphi
\sin^2\theta \cos\phi_1 d\theta \wedge d\phi_2
-\frac{np g_s^2}{(m^2 + n^2 g_s^2)}(1-H'^{-1})
dx^4 \wedge dx^5\br
A^{(1)} &=& - \frac{n}{m}(H''^{-1}-1) dx^3\br
A^{(3)} &=& \frac{2n}{\sqrt{m^2 + n^2 g_s^2}}Q_5\cos\varphi
\sin^2\theta \cos\phi_1 dx^3 \wedge d\theta \wedge d\phi_2\br
&& +\frac{mp}{(m^2 + n^2 g_s^2)}(1-H'^{-1})
dx^3 \wedge dx^4 \wedge dx^5\ - \frac{\sin\varphi}{g_s} H^{-1} dx^0 \wedge 
dx^1 \wedge dx^2
\lll{28}
\eea
It can be easily checked from the above solution that for $p=0$, it reduces
to (NS5, D4) solution given in \eqn{6} and for $n=0$, it reduces to (NS5, D2)
solution given in \eqn{9}. Also for $m=0$, the above solution reduces to
(D4, D2) solution (with additional isometries in $x^3$ direction) constructed 
in \cite{breckmm,cosp}.
Since the above bound state belongs to type IIA theory, we apply the T-duality
map from type IIA to type IIB fields in \eqn{7} along $x^2$ coordinate. Thus we
obtain the (NS5, D3, D1) bound state where D3-branes lie along $x^1,\,x^4,\,
x^5$ coordinates and D-strings lie along $x^1$ coordinate.

\noindent{\bf \underline{(NS5, D3, D1) solution:}}
\bea
ds^2 &=& 
H^{1/2} H''^{1/2}\left[H^{-1}\left(-dx_0^2 + dx_1^2\right) + H''^{-1}
\left(dx_2^2 + dx_3^2\right) 
+ H'^{-1}\left(dx_4^2 + dx_5^2\right) + dr^2 + r^2 d\Omega_3^2\right]\br
e^{\phi_b} &=& g_s H^{1/2}H'^{-1/2} H''^{1/2}\br
B^{(b)} &=& \frac{2m}{\sqrt{m^2 + n^2 g_s^2}}Q_5\cos\varphi
\sin^2\theta \cos\phi_1 d\theta \wedge d\phi_2
-\frac{np g_s^2}{(m^2 + n^2 g_s^2)}(1-H'^{-1})
dx^4 \wedge dx^5\br
\chi &=& 0\br
A^{(2)} &=& - \frac{n}{m}(H''^{-1}-1)dx^2 \wedge dx^3 - \frac{\sin\varphi}{g_s}
H^{-1} dx^0 \wedge dx^1\br
A^{(4)} &=& - \frac{n (H''^{-1} + 1)}{\sqrt{m^2 + n^2 g_s^2}}Q_5\cos\varphi
\sin^2\theta \cos\phi_1 dx^2 \wedge dx^3 \wedge d\theta \wedge d\phi_2\br
&& -\frac{mp}{(m^2 + n^2 g_s^2)}(1-H'^{-1}) dx^2 \wedge
dx^3 \wedge dx^4 \wedge dx^5\br
&& + \frac{1}{2} \frac{n^2 p g_s^2}{m(m^2 + n^2 g_s^2)}(H''^{-1} - 1)
(1 - H'^{-1}) dx^2 \wedge dx^3 \wedge dx^4 \wedge dx^5 
\lll{29} 
\eea
It can be checked from above that for $n = 0$, the solution reduces to
(NS5, D1) solution given in eq.\eqn{10}. However, for $p = 0$, it does not
quite give the (NS5, D3) bound state in eq.\eqn{8}. The reason for this is
that, we have not included the other components of $A^{(4)}$ in eq.\eqn{29}
obtained by the self-duality condition on the corresponding field strength.
The field strength associated with $A^{(4)}$ is given below,
\bea
F^{(5)} &=& -\frac{2n}{\sqrt{m^2 + n^2 g_s^2}}H''^{-1}Q_5\cos\varphi\sin^2
\theta\sin\phi_1 dx^2 \wedge dx^3 \wedge d\theta \wedge d\phi_1 \wedge 
d\phi_2\br
&& + \frac{mp}{m^2 + n^2 g_s^2} dH'^{-1} \wedge dx^2 \wedge dx^3 \wedge
dx^4 \wedge dx^5\br
&& - \frac{n^2 p g_s^2}{m(m^2 + n^2 g_s^2)}(H''^{-1} - 1) dH'^{-1} \wedge
dx^2 \wedge dx^3 \wedge dx^4 \wedge dx^5\br
&& + \frac{m\sin\varphi\cos\varphi}{g_s\sqrt{m^2 + n^2 g_s^2}}Q_5 \sin^2\theta
\cos\phi_1 dH^{-1} \wedge dx^0 \wedge dx^1 \wedge d\theta \wedge d\phi_2\br
&& - \frac{m\sin\varphi\cos\varphi}{g_s\sqrt{m^2 + n^2 g_s^2}}Q_5 \sin^2\theta
\sin\phi_1 H^{-1} dx^0 \wedge dx^1 \wedge d\theta \wedge d\phi_1 \wedge 
d\phi_2\br
&& - \frac{1}{2}\frac{npg_s^2}{m^2 + n^2 g_s^2} \frac{\sin\varphi}{g_s}
(1-H'^{-1}) dH^{-1} \wedge dx^0 \wedge dx^1 \wedge dx^4 \wedge dx^5\br
&& - \frac{1}{2}\frac{npg_s^2}{m^2 + n^2 g_s^2} \frac{\sin\varphi}{g_s}
H^{-1} dH'^{-1} \wedge dx^0 \wedge dx^1 \wedge dx^4 \wedge dx^5
\lll{30}
\eea
Here, note that if we set $p = 0$, all the terms in $F^{(5)}$ except the first
term vanish. By simply taking the Hodge duality on this term correctly
reproduces the form of $A^{(4)}$ in (NS5, D3) solution in eq.\eqn{8}. We
also note that by taking $m = 0$, above solution again does not quite reduce
to (D3, D1) solution (with additional isometries in $x^2$, $x^3$ directions)
obtained in \cite{breckmm,cosp,russot}. The reason again is that we have not 
included the terms
obtained by Hodge duality in $A^{(4)}$. This is required because D3-branes are
self-dual. However, for complicated bound state system involving D3, like the
case we are considering, it is not clear how to write the Hodge dual terms
in the gauge field since they produce non-local terms in general. But since we
know the explicit form of the field strength we can verify that for the 
special case of $m = 0$, the Hodge duality correctly reproduces the required
terms in $A^{(4)}$. Thus we recover the (D3, D1) solution from above by setting
$m = 0$.

The (NS5, D3, D1) solution constructed above belongs to type IIB theory. So, 
we use the T-duality map given in \eqn{4} along $x^1$-coordinate to obtain
(NS5, D2, D0) solution. Here NS5-branes lie along $x^1,\,x^2,\cdots,x^5$
directions and D2-branes along $x^4,\,x^5$-directions.

\noindent{\bf \underline{(NS5, D2, D0) solution:}}
\bea
ds^2 &=& 
H^{1/2} H''^{1/2}\left[-H^{-1} dx_0^2 + H''^{-1}
\left(dx_1^2 + dx_2^2 + dx_3^2\right) 
+ H'^{-1}\left(dx_4^2 + dx_5^2\right) + dr^2 + r^2 d\Omega_3^2\right]\br
e^{\phi_a} &=& g_s H^{3/4}H'^{-1/2} H''^{1/4}\br
B^{(a)} &=& \frac{2m}{\sqrt{m^2 + n^2 g_s^2}}Q_5\cos\varphi
\sin^2\theta \cos\phi_1 d\theta \wedge d\phi_2
-\frac{np g_s^2}{(m^2 + n^2 g_s^2)}(1-H'^{-1})
dx^4 \wedge dx^5\br
A^{(1)} &=& \frac{\sin\varphi}{g_s} H^{-1} dx^0\br
A^{(3)} &=& - \frac{n}{m}(H''^{-1}-1)dx^1 \wedge dx^2 \wedge dx^3 + 
\frac{m\sin\varphi\cos\varphi}{g_s\sqrt{m^2 + n^2 g_s^2}} Q_5 H^{-1} 
\sin^2\theta \cos\phi_1 dx^0 \wedge d\theta \wedge d\phi_2\br
&& -\frac{1}{2}\frac{npg_s}{(m^2 + n^2 g_s^2)} \sin\varphi H^{-1}(1-H'^{-1})
dx^0 \wedge dx^4 \wedge dx^5 
\lll{31} 
\eea
Again for $n = 0$, we recover the (NS5, D0) solution given in eq.\eqn{11}.
Similarly, by setting $m = 0$, we recover the (D2, D0) bound state solution
(with additional isometries in $x^1$, $x^2$, $x^3$ directions) obtained in
\cite{breckmm,cosp,russot}. However, for $p = 0$, we can only recover 
(NS5, D2) solution given
in eq.\eqn{9} if we include the Hodge dual terms in $A^{(4)}$ of (NS5, D3, D1)
solution in eq.\eqn{29} for this special case as mentioned before.

The ADM mass and the tension of  (NS5, D$(p+2)$, D$p$) bound states can be
obtained by a further generalization of the mass formula given in 
\cite{lurtwo}. We find that it is given by,
\be
T_{(m,n,p)} = \frac{1}{g_s^2}\sqrt{(p^2 + n^2)g_s^2 + m^2}\,\frac{1}{(2\pi)^5
\alpha'^3}
\lll{t2}
\ee
Again it is proportional to the charge $Q_5$ given in \eqn{27}. It can be 
easily verified from \eqn{t2} that when any two of the integers $m$, $n$, $p$
are relatively prime the (NS5, D$(p+2)$, D$p$) form non-threshold bound states.

We would like to point out that one can make further SL(2, {\bf Z})
transformation on (NS5, D3, D1) to construct the most general bound state
involving NS5-branes and lower D$p$-branes of the form (NS5, D5, D3, D1, F) of
type IIB string theory. Furthermore, if we take T-duality on this state along
$x^1$-direction, we obtain type IIA bound state of the form (NS5, D4, D2, D0,
W), where `W' denotes the waves. However, it is not clear how to obtain a
similar state involving F-string in this case as given in \cite{alioz}.

\section{OD$p$-limit}

It has been shown in \cite{gopmss} that there exist a series of new 
six-dimensional theories which are nothing but the decoupled theories of
NS5-branes in the presence of a critical RR $(p+1)$-form gauge field whose
excitations include light open D$p$-branes known as OD$p$ theories. The
dual supergravity solutions of these theories are the particular decoupling
limit of the (NS5, D$p$) bound state configurations constructed in section 2.
Although these supergravity solutions have already been given in \cite{alior},
we here briefly discuss these solutions since our solutions differ from those
in \cite{alior}. The OD$p$-decoupling limit is defined as follows: 
\be
\cos\varphi = \epsilon \to 0
\lll{32}
\ee
keeping the following quantities fixed
\be
\alpha'_{{\rm eff}} = \frac{\alpha'}{\epsilon}, \quad u = \frac{r}{\epsilon
\alpha'_{{\rm eff}}}, \quad G_{o(p)}^2 = \epsilon^{(p-3)/2}g_s,
\quad Q_5 = \alpha'_{{\rm eff}}q
\lll{33}
\ee
The harmonic functions take the forms
\be
H = \frac{1}{\epsilon^2 a^2 u^2}, \qquad H' = \frac{h}{a^2u^2}
\lll{34}
\ee
where $h = 1 + a^2 u^2$, with $a^2 = \frac{\alpha'_{{\rm eff}}}{q}$.
The metric, dilaton and the NSNS 2-form are:
\bea
ds^2 &=& \alpha' h^{1/2}\left[-d\tilde{x}_0^2 + \sum_{i=1}^p d\tilde{x}_i^2
+ h^{-1}\sum_{j=p+1}^5 d\tilde{x}_j^2 + \frac{q}{u^2}\left(du^2 + u^2 
d\Omega_3^2\right)\right]\br
e^{\phi_{(a,b)}} &=& G_{o(p)}^2 \frac{h^{(p-1)/4}}{au}\br
B_{\theta\phi_2}^{(a,b)} &=& 2\alpha' q \sin^2\theta \cos\phi_1
\lll{35}
\eea
where $\tilde{x}_{0,\cdots,p} = \frac{1}{\sqrt{\alpha'_{{\rm eff}}}} 
x_{0,\cdots,p}$,
$\tilde{x}_{p+1,\cdots,5} = \frac{\sqrt{\alpha'_{{\rm eff}}}}{\alpha'}
x_{p+1,\cdots,5} = {\rm fixed}$. The quantization condition is
\be
\frac{p}{q} = \frac{\tan\varphi}{g_s} = \frac{1}{G_{o(p)}^2 \epsilon^{(5-p)/2}}
\lll{36}
\ee
The RR gauge fields take the following forms for different values of $p$,
\bea
p &=& 0, \qquad A_0^{(1)} = \frac{\alpha'^{1/2}}{G_{o(0)}^2}(au)^2,
\quad A_{0\theta\phi_2}^{(3)} = \frac{\alpha'^{3/2} q}{G_{o(0)}^2}(au)^2
\sin^2\theta \cos\phi_1\br
p &=& 1, \qquad \chi = 0,\quad A_{01}^{(2)} = 
-\frac{\alpha'}{G_{o(1)}^2}(au)^2,
\quad A_{2345}^{(4)} = \frac{\alpha'^2}{G_{o(1)}^2}\frac{(au)^2}{h}\br
p &=& 2, \qquad A^{(1)} = 0,\quad A_{012}^{(3)} = 
-\frac{\alpha'^{3/2}}{G_{o(2)}^2}(au)^2,
\quad A_{345}^{(3)} = -\frac{\alpha'^{3/2}}{G_{o(2)}^2}\frac{(au)^2}{h}\br
p &=& 3, \qquad \chi = 0,\quad A_{45}^{(2)} = 
-\frac{\alpha'}{G_{o(3)}^2}\frac{(au)^2}{h},
\quad A_{45\theta\phi_2}^{(4)} = \frac{\alpha'^2 q}{G_{o(3)}^2}(1+h^{-1})
\sin^2\theta\cos\phi_1,\br 
&&\qquad\qquad\qquad\quad A_{0123}^{(4)} = -\frac{\alpha'^2}{G_{o(3)}^2}
(au)^2\br
p &=& 4, \qquad A_5^{(1)} =  
-\frac{\alpha'^{1/2}}{G_{o(4)}^2}\frac{(au)^2}{h},
\quad A_{5\theta\phi_2}^{(3)} = -\frac{2q\alpha'^{3/2}}{G_{o(4)}^2}
\sin^2\theta\cos\phi_1
\br
p &=& 5, \qquad \chi =  
-\frac{1}{G_{o(5)}^2}\frac{1}{h},
\quad A_{\theta\phi_2}^{(2)} = -2p\alpha'
\sin^2\theta\cos\phi_1
\lll{37}
\eea
Note that in writing down the above solutions we have thrown away some constant
pieces in the gauge fields $A^{(4)}_{2345}$, $A_{345}^{(3)}$, $A_{45}^{(2)}$
and $A_5^{(1)}$ in $p=1,\,2,\,3,\,4$ respectively.  
For $p=3$ we have added the dual of 
the gauge field
$A_{0123}^{(4)}$ and this is crucial to have the correct form of (NS5, D4)
and (NS5, D5) solutions written in \eqn{1} and \eqn{6} if we start from 
(NS5, D1)
solution and apply T-duality to obtain them as done in \cite{alior}.  
However, the metric and the dilaton which essentially 
determine the behavior of the theory at various regimes of the 
energy parameter
$u$ have the same form as in \cite{alior} and therefore, 
the conclusions remain the same.
We will here briefly discuss the OD$p$-limits for different values of $p$.

It has been noted in refs.\cite{ali,har} that OD$p$-limit 
(at least for $p=1,\,2$), is 
different from the little string theory limit $g_s \to 0$ and $\alpha'=$fixed.
However, it has been clarified later in \cite{alior} (also in \cite{har}), 
that for $p \leq 2$, 
one can 
take a decoupling limit different from \eqn{33} and resembles the little string 
theory limit as follows\footnote{For $p=1$ and $p=2$, these limits have been
referred to in \cite{har} as the open brane little string theory (OBLST) 
limits. The solutions (37)--(39) in these cases are the supergravity dual
descriptions of (1, 1) OBLST and (0, 2) OBLST respectively. We would like to
thank T. Harmark for pointing this out to us.}:
\be
\cos\varphi=\epsilon\to 0, \quad g_s=\epsilon^{(3-p)/2} G_{o(p)}^2 \to 0,
\quad u=\frac{r}{\sqrt{\alpha'}\sqrt{\epsilon} G_{o(p)}^{2/(3-p)}}={\rm fixed},
\quad \alpha'={\rm fixed}
\lll{38}
\ee
The supergravity solution reduces precisely to the same form as in OD$p$ given
in eqs.\eqn{35}--\eqn{37} with the coordinate scaling
\be
\tilde{x}_{0,\cdots,p}=\sqrt{\frac{\epsilon}{\alpha'}} x_{0,\cdots,p}
= {\rm fixed},
\qquad \tilde{x}_{p+1,\cdots,5}=\frac{1}{\sqrt{\alpha'\epsilon}} x_{p+1,\cdots,
5} = {\rm fixed}
\lll{39}
\ee
but now $\alpha'={\rm fixed}$ (as opposed to OD$p$-limit where $\alpha'\to 0$)
and $a^2=G_{o(p)}^{4/(3-p)}/q$. Indeed, we note from eq.\eqn{35} that for
$au \ll 1$, we recover (5+1)-dimensional Lorentz invariance and the 
supergravity solutions in this case reduce to those of little string theories
given in \cite{imalsy,ahabks}.

Now coming back to OD$p$ theories, we note that the curvature of the metric
in eq.\eqn{35} measured in units of $\alpha'$ i.e. 
$\alpha'{\cal R} \sim 1/q$ and
so, supergravity solution can be trusted for large enough $q$ (no. of 
NS5-branes). When $au \ll 1$, i.e. in the IR region the supergravity solution
is valid only if $au \gg G_{o(p)}^2$, where $e^{\phi_{(a,b)}}$ remains small.
In this case $G_{o(p)}^2 \ll 1$. However, this condition is not satisfied in
the extreme IR region, where $e^{\phi_{(a,b)}}$ is large. In that case, we
have to either go to the S-dual frame for type IIB theory or lift the solution
to M-theory for type IIA theory to have valid supergravity description. The
OD$p$-theories in this case flow to (5+1)-dimensional SYM theory for $p$ = odd
and for $p$ = even, they flow to (0, 2) superconformal field theory.

For $au \gg 1$, i.e. in the UV region, the supergravity solution is valid if
$(au)^{(3-p)/2} \gg G_{o(p)}^2$. So, for $p=0,\,1,\,2$ we have $au \gg
G_{o(0)}^{4/3}$, $au \gg G_{o(1)}^2$ and $au \gg G_{o(2)}^4$ respectively.
In the extreme UV region, these conditions are always satisfied and we have
good supergravity description. The form of the metric in these cases reduce
to those of ordinary D$p$-branes with additional isometries in $(5-p)$
directions. The reason for this can be understood from the quantization 
relation eq.\eqn{36}, where we note that the D$p$-branes dominate over 
NS5-branes.

For $p=3$, the string coupling $e^{\phi_b} = G_{o(3)}^2 = {\rm fixed}$. So,
when $G_{o(3)}^2 \ll 1$, the metric reduces to that of ordinary D3-branes with 
additional isometries in 4, 5 directions. This can also be understood from
the quantization relation \eqn{36}. When $G_{o(3)}^2 \gg 1$, we have 
to go to the
S-dual frame and (NS5, D3) solution becomes (D5, D3) system and thus the 
strongly coupled OD3 theory become equivalent to six-dimensional NCYM with
noncommutativity parameter $\theta = 2\pi \alpha'_{{\rm eff}} G_{o(3)}^2$ and
the coupling $g_{YM}^2 = (2\pi)^3 \alpha'_{{\rm eff}}$.

For $p=4$, the string coupling $e^{\phi_a} = G_{o(4)}^2 (au)^{1/2}$. So, the
supergravity solution is valid if $au \ll G_{o(4)}^{-4}$. In the extreme UV
region this condition is not satisfied and therefore, the supergravity 
description would have to be given by lifting the solution to M-theory. The
supergravity solution is then given by the two intersecting M5-branes along
1,2,3,4 directions.

For $p=5$, the string coupling $e^{\phi_b} = G_{o(5)}^2 au$. So, in order
to have valid supergravity description $au \ll G_{o(5)}^{-2}$. This again
is not satisfied if we are in the extreme UV region. In this case, we have
to go to the S-dual frame. Note from \eqn{37} that in UV, $\chi \to 0$ (This
is because in our solution of (NS5, D5) given in eq.\eqn{1}, the asymptotic
value of $\chi$ vanishes) and therefore, the string coupling in the S-dual
frame is given as $e^{\phi'_b} = (G_{o(5)}^2 au)^{-1}$ and remains small. The 
metric $ds'^2 = e^{-\phi_b} g_s ds^2$ and the dilaton as given above then 
reduces to those of little string theory with $g_s' \to 0$ and $\alpha'_{{\rm
eff}}$ = fixed. We note from eq.\eqn{36} that $G_{o(5)}^2 = q/p$ is quantized
and the axion in \eqn{37} reduces to a rational number in IR, whereas it 
vanishes in the UV. Here, our conclusion differs from ref.\cite{alior} and 
this is 
because we have taken the asymptotic value of the axion to be zero, whereas in
ref.\cite{alior}, it was taken to be constant.

\vskip .5cm

\noindent{\bf Note added:}

After submission of this paper to the net we were informed by M. Cederwall,
U. Gran, M. Nielsen and B. E. W. Nilsson that some of the solutions 
(for type IIB) constructed in this paper were also considered in \cite{cgnn}
from a different approach.

\section*{Acknowledgements}

We would like to thank Jianxin Lu for discussions, very useful suggestions and
a critical reading of the manuscript. We would also like to thank M. 
Alishahiha, T. Harmark for e-mail correspondences, M. Costa for informing us
about \cite{cosp} and M. Cederwall, U. Gran, M. Nielsen and B. E. W. Nilsson
for informing us about \cite{cgnn}.

\end{document}